\documentclass[pre,twocolumn,showpacs]{revtex4}
\newcommand{\figurewidth}{1.\columnwidth}

\usepackage{graphics}
\usepackage{amsmath}
\usepackage{amssymb}
\usepackage{dcolumn}

\begin{document}

\title{Growth by Random Walker Sampling, and Scaling of the Dielectric
Breakdown Model}

\author{Ell\'{a}k Somfai}
\email{ellak@lorentz.leidenuniv.nl}
\altaffiliation[Present address: ]{Universiteit Leiden, Instituut-Lorentz,
PO Box 9506, 2300 RA Leiden, The Netherlands}
\affiliation{Department of Physics, University of Warwick, Coventry CV4
7AL, United Kingdom}

\author{Nicholas R. Goold}
\email{N.R.Goold@warwick.ac.uk}
\affiliation{Department of Physics, University of Warwick, Coventry CV4
7AL, United Kingdom}
\author{Robin C. Ball}
\email{r.c.ball@warwick.ac.uk}
\affiliation{Department of Physics,
University of Warwick, Coventry CV4
7AL, United Kingdom}

\author{Jason P. DeVita}
\email{jdevita@umich.edu}
\affiliation{Michigan Center for Theoretical Physics, Department of
Physics, University of Michigan, Ann Arbor, Michigan, 48109-1120}

\author{Leonard M. Sander}
\email{lsander@umich.edu}
\affiliation{Michigan Center for Theoretical Physics, Department of
Physics, University of Michigan, Ann Arbor, Michigan, 48109-1120}

\date{\today{}}
\begin{abstract}

Random walkers absorbing on a boundary sample the Harmonic Measure
linearly and independently: we discuss how the recurrence times between
impacts enable non-linear moments of the measure to be estimated.
From this we derive a new technique to simulate Dielectric Breakdown
Model growth which is governed nonlinearly by the Harmonic Measure.
Recurrence times are shown to be accurate and effective in probing
the multifractal growth measure of diffusion limited aggregation.
For the Dielectric Breakdown Model our new technique grows large clusters
efficiently and we are led to significantly revise earlier exponent
estimates. Previous results by two conformal mapping techniques were
less converged than expected, and in particular a recent theoretical
suggestion of superuniversality is firmly refuted.
\end{abstract}
\maketitle

\section{Introduction}

The steady state distribution of random walking particles is governed by
Laplace's Equation. As a result Witten and Sander's model of Diffusion
Limited Aggregation (DLA) \cite{witten81}, in which a cluster grows by
irreversible
accretion of dilute diffusing material, has been of double interest: it
is readily simulated out to huge cluster sizes \cite{ball85, tolman89}, whilst
at the same time
the governance of its growth by Laplace's equation renders it a landmark
mathematical challenge to analyse. The connection to a Laplacian field
also underpins the breadth of application of the DLA model, to problems
such as viscous fingering in porous media \cite{nittmann85} and
electrodeposition \cite{matsushita84, brady84}. This
interest has all been abetted by controversy as to whether the distribution
of cluster shapes conforms to simple fractal scaling (see \cite{somfai99} and references therein), and interest in the multifractal scaling of the growth measure \cite{amitrano89, ball90blunt, jensen03, halsey86prl}.

Physical analogies and the mathematical connections have led to interest in
other models where the growth is governed \emph{non-linearly} by a Laplacian
field. In particular Niemeyer, Pietronero and Wiesmann introduced
the family of Dielectric Breakdown Models (DBM) \cite{niemeyer84} where

\begin{equation}
v_{n}\propto \left| \partial _{n}\phi \right| ^{\eta },\qquad \nabla
^{2}\phi =0,\text{\qquad }\phi _{\text{interface}}\approx 0  \label{growth}
\end{equation}
and $\eta $ is (for interest) a positive parameter. Interest in this is
further prompted by Ball and Somfai's proposal \cite{ball02prl,ball03pre}
that growth proportional to field with
non-trivial spatial cutoff maps onto simple DBM but with $\eta\not=1$. It is
important that diffusion controlled growth which is locally limited by the
capillary energy associated with surface ramification is a case in point.
Computationally these non-linear models have been a challenge, as random
walkers only directly sample the harmonic measure linearly, realising only
the $\eta =1$ case.

First in this paper we show that random walkers can be exploited to
sample non-linear moments of the harmonic measure.  In this way we obtain
results for the active portion of the multifractal spectrum of DLA far
beyond existing results.  The key strength of these methods is that no
explicit solving of the Laplace equation is involved.

We then exploit this to establish a new method of growing DBM clusters by
random walker accretion.  This also entails adopting the noise reduction
strategies lately introduced in Ref.~\cite{ball02pre},  and enables us to
explore the DBM class out to unprecedentedly large clusters. We show that in
two dimensions this largely resolves how the exponents of the DBM model depend
on $\eta$: in particular superuniversality of the tip scaling exponent
alpha is
strongly refuted,  in favour of a continuous variation of exponents which
also confirms the hypothesised  upper critical value \cite{hastings01pre,
hastings01prl, halsey02, sanchez93} $\eta_c=4$.

Our walker-DBM results are supported by extensive computations using
established
iterative conformal mapping methods due to Hastings and Levitov (HL)
\cite{hastings98} and also by direct integration of the
Shraiman-Bensimon Equations \cite{shraiman84} exploiting the
mappings of Ball and Somfai \cite{ball02prl,ball03pre}.
For exponents, all three agree within
statistical errors.  Below the level of the errors there is a systematic difference
between walker-DBM and HL,  and separate results for the relative penetration
depth suggest that it is the walker-DBM clusters which are more converged to
asymptotic behaviour.  Unlike HL and our Shraiman-Bensimon integrations, the
new walker-DBM technique is not limited to two dimensions and so the way
forward appears open to a full exploration of the DBM class in three
dimensions.

\section{Random Walker Sampling}
\label{sec:rwsampling}

We consider first the problem of sampling the harmonic measure of an
equipotential surface.  The harmonic measure is given on the surface by
$\frac{d\mu}{d s}=\frac{\partial\phi}{\partial n}$ where
outside the surface the scalar field $\phi$ obeys a Poisson equation
$-\nabla^{2}\phi=S(x)$ with $S(x)$ the source density; typically we will be
interested in cases where the source is concentrated at
points, particularly
$\infty$.  We can sample this measure by introducing random walkers at the
source points,  and tracing their Brownian trajectories to the point of
first encounter with the surface,  whereupon any given walker is discarded.
The points of first encounter uniformly sample the measure $\mu$.

By firing a large enough sample of random walkers and collecting frequency
counts of their hit distribution, one can build up an approximation to the
entire harmonic measure.  This has been successfully used by Somfai \emph{et
al.}
\cite{somfai99}. However such methods are expensive and give differing
quality estimates across the support.

Here we focus on recurrence times defined as follows.  We first divide the
support into (many) small partitions (hereafter termed ``sites'') for each of
which we aim to estimate the corresponding hit probability $\mu$. We then fire
(independent) walkers sequentially at the surface and when each walker hits,
the number of walkers fired since the previous hit on that site we will
call the ``age'' $a$,  and this provides a simple estimate of the hit
probability of that site $\mu_1=1/a$.
This is a standard way to estimate frequency of uncorrelated events,
by their recurrence time.  The probability distribution of the estimator $\mu_1$, given the true underlying value
$\mu$ for that site,
is given by simple Poisson statistics as
\[
p_{1}(a|\mu)=\mu e^{-\mu a},
\]
assuming that $\mu\ll 1$ so that we can approximate age as a continuous
variable.  We can generate more reliable estimates by using the age $a_{k}$
since the $k$'th previous hit,  which is distributed according to
\begin{equation}
p_{k}(a_{k}|\mu)=\mu e^{-\mu a_{k}}(\mu a_{k})^{k-1}/(k-1)!  \label{pk},
\end{equation}
in terms of which our estimate is $\mu_k=k/a_{k}$.

In applications considered below we will be interested in calculating
moments of the measure, $M_{q}=\sum \mu^{q}$, where the summation is over
the sites over which we partitioned the support.  This can be interpreted
as $M_{q}=\left\langle \mu^{q-1}\right\rangle $ where the average is
weighted by the measure $\mu$ itself which in turn is sampled by where
random walkers hit.  At each impact we can use the site age in our estimate
for the factors $\mu^{q-1}$,  leading to the estimate
\[
M_{qk}=A(k,q)\left\langle (a_{k}/k)^{1-q}\right\rangle,
\]
where the average is now over all random walkers hitting the surface,  we
use $k$'th ages, and $A(k,q)$ is a trivial numerical front factor.

Random walkers being cheap,  our main concern is when this estimate
converges.  Using the distribution (\ref{pk}) one can readily check that
$M_{qk}$ converges to $M_{q}$ provided  $k>q-1$, and we use the appropriate prefactor
 $A(k,q)=\frac{k^{1-q}(k-1)!}{(k-q)!}$.
Thus we can compute moments of finite degree simply by
using high enough order ages.  In practice we will see that for problems of
interest the most significant limitation comes not at high $q$ but rather
at highly negative $q$,  for which random walkers give an inefficient
sample of the relevant parts of the measure.

\section{Moments under Growth}

The simple ideas above become rather useful when extended to compute moments
of the measure as a surface grows.  The harmonic measure of Diffusion
Limited Aggregation provides a well studied (but not entirely resolved)
test case.  For a cluster of $N$ added particles the conventional
multifractal scaling would lead to $M_{q}\sim N^{-\tau(q)/D}$,  where $D$
is the fractal dimension relating radius $R$ to $N$ through $N\sim R^{D}$.
Summing these moments over $N$ with weight $N^{t-1}$gives us a partition
function $Z(q,t)$ which we can estimate (ignoring numerical prefactors) as
\[
Z(q,t,N)=\sum \limits_{n=1}^{N}a^{1-q}n^{t-1},
\]
where the sum is now over the particles used to grow the cluster and the
corresponding ages of the sites where they hit.  Following the spirit of
how Halsey \emph{et al.} \cite{halsey86pra} generalised the identification of
multifractal spectrum,  we can now identify that $\tau(q)/D$ separates the values
$t<\tau(q)/D$ for which $Z(q,t,N)\rightarrow\infty$ as $N\rightarrow\infty$
from the values $t>\tau(q)/D$ for which $Z(q,t,N)\rightarrow0$.

The above definition does not restrict the behaviour of $Z(q,t,N)$ on the
locus $t=\tau(q)/D$ but the simple expectation is of a logarithmic
divergence with $N$.  Then a numerical strategy is to choose $t$ such that
$Z(q,t,N)-Z(q,t,\epsilon N)=\sum \limits_{n=\epsilon N+1}^{N}a^{1-q}n^{t-1}$
becomes independent of $N$ as $N$  becomes large with fixed $\epsilon$.

To obtain results at $q\geq 2$ we have to use higher order ages.  The first
order age is naturally thought of as the age of the parent to a given new
site and a corresponding estimate of $a_{2}$ is given by the age of its
grandparent (the parent of its parent) and so on.  It is of course a
concern that the more such generations one looks back for the age,  the
more the cluster geometry will have evolved in the meantime:  we will see
that overall cluster screening cuts in below a threshold value of $q$ in a
well understood way.  To alleviate the worry above threshold we have
studied noise reduced clusters where each particle deposited advances the
local geometry by only a small fraction $A$ of one notional particle size,
so that we can be confident that the local geometry does not evolve
significantly on looking back of order $1/A$ generations.

\begin{figure}
\resizebox{\figurewidth}{!}{\includegraphics*{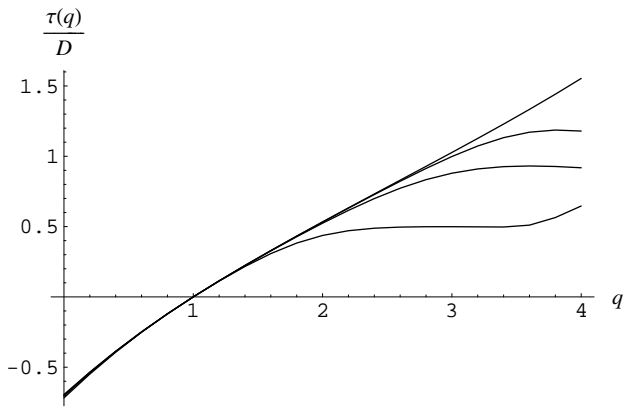}}
\caption{\label{fig:tauvsq}
The moment spectrum $\tau (q)$ of the harmonic measure obtained over the
growth of a single DLA cluster by imposing the condition
$Z(q,t,N)-Z(q,t,N/4)=Z(q,t,N/4)-Z(q,t,N/16)$ to yield $t=\tau(q)/D$.  The
respective curves used ages of order $k=1,2,4,8$ and can be seen to break
away around $q=k+1$ as expected from our discussion in section
\ref{sec:rwsampling}.  Note also close agreement with the electrostatic
scaling law \cite{halsey87} that $\tau(3)/D=1$.
These results come from a single off-lattice DLA cluster of $N=10^{6}$
particles grown with noise reduction factor \cite{ball02pre}
$A=0.1$.
}
\end{figure}

Figure~\ref{fig:tauvsq} shows the moment spectrum $\tau (q)$ resulting from imposing the
condition $Z(q,t,N)-Z(q,t,\epsilon N)=Z(q,t,\epsilon N)-Z(q,t,\epsilon ^{2}N)
$ on a {\it single} off-lattice DLA cluster. As expected
from our discussion in section \ref{sec:rwsampling},  the results using ages
of order $k$ break away at
$q=k+1$.  An important check is given by the measured value $\tau (3)/D=1.01
$ compared to the prediction of unity from the Electrostatic Scaling Law.

\begin{figure*}
\resizebox{0.9 \textwidth}{!}{\includegraphics*{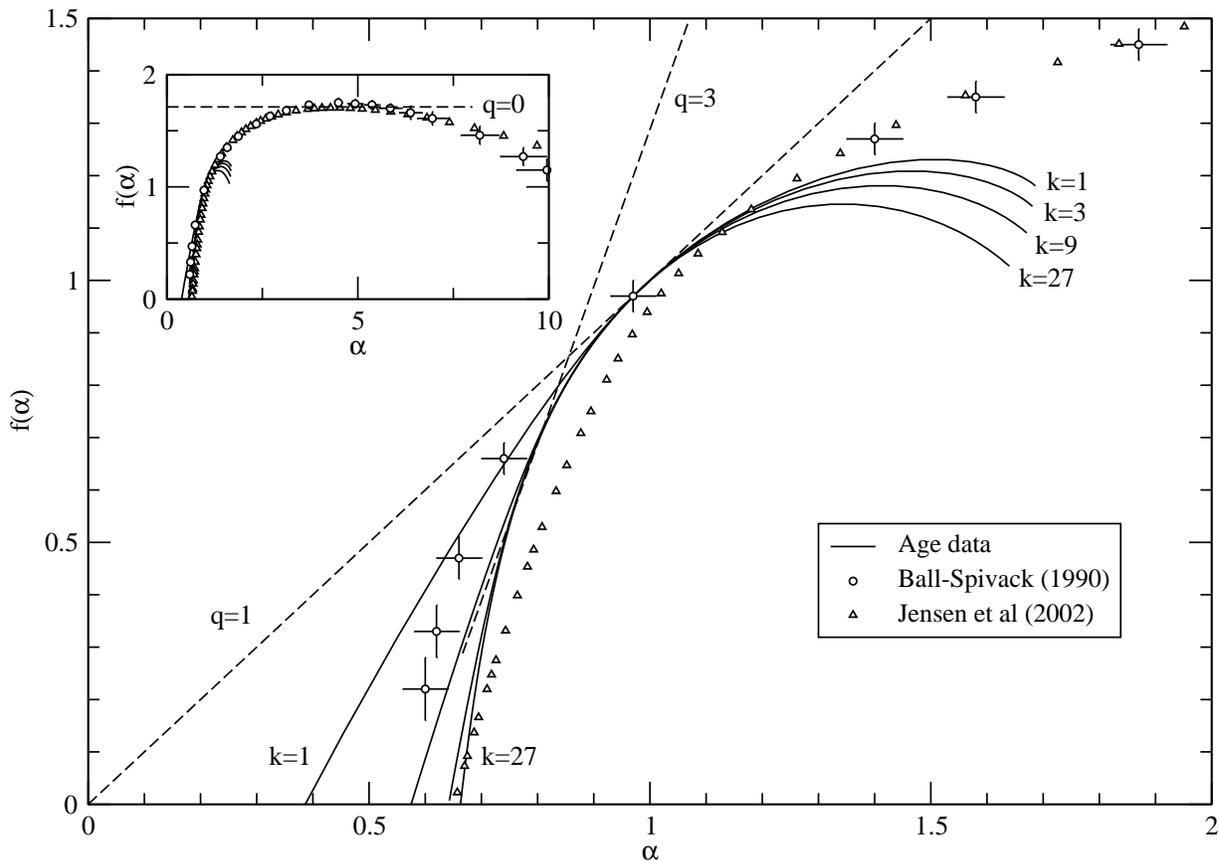}}
\caption{\label{fig:mfspec}
Multifractal spectrum of [the harmonic measure of] DLA obtained using ages
of order
$k=1,3,9,27$. Also shown are earlier results from Ball and Spivack
\cite{ball90} and Jensen \emph{et al.} \cite{jensen02}.  Note how much better the present data agrees with
the  tangent lines
representing Halsey's electrostatic scaling law (for $q=3$) and Makarov's
theorem (for $q=1$). The inset shows how the older data extends to higher
$\alpha$, which our method cannot probe.
}
\end{figure*}

For more definitive results we have studied a sample of $10$ clusters out to
$N=10^{7}$ using $A=0.1$.  Even allowing for the noise reduction which
means that of order $1/A=10$ particles were deposited per local unit of
advance in the growth,  it is unprecedented to probe the harmonic measure
out to such large scales.  Because of the volume of age data,  it was
cumbersome to seek stationarity of $Z(q,t,N)-Z(q,t,\epsilon N)$ and instead
we extracted $t(q)$ from the simple scaling expectation that $Z(q,t^{\prime
},N)-Z(q,t^{\prime },\epsilon N)\propto N^{t^{\prime }-t(q)}$ where we
simply chose $t^{\prime }$ to assure reasonably uniform weighting of the
data. We then took $\tau (q)=D t(q)$ using $D=1.71$,  where any error in
this value is not significant to the accuracy of $t(q)$.  The resulting
$f(\alpha )$ spectrum is shown in Fig.~\ref{fig:mfspec} for ages of order $k=1,3,9,27$
and
compared to one earlier report of the $f(\alpha )$ spectrum which is
distinguished by carrying full error bars \cite{ball90} and one more recent spectrum \cite{jensen02} which
claimed better convergence than all previous.

The success of our data lies in the region $q\geq 1$ corresponding to
$\alpha \leq 1$,  where it passes two important tests by meeting the
(dashed) tangent lines corresponding to Makarov's Theorem\cite{makarov85} at $q=1$ and (for
the higher $k$ as appropriate) Halsey's electrostatic scaling law\cite{halsey87} at $q=3,$
and in these respects it clearly improves on the earlier published data.

A significant limitation of the $f(\alpha )$ spectrum obtained is that it
substantially undershoots at large $\alpha $, but this is qualitatively
consistent with expectations.  It is obvious that we cannot expect to probe
probabilities to hit a site which are smaller than $1/N$ as these are
unlikely to be sampled by the $N$ walkers used to grow the cluster.  Thus we
certainly cannot probe the spectrum for $\alpha >D$.

A more careful argument suggests that in principle we should (in the limit
of large enough $N$) be able to probe all the way up to $\alpha =D$,  as
follows.
Consider a typical site $X$ ``born'' with hit probability $R^{-\alpha }$,
$\alpha <D$.  It is expected that the probability for growth within distance
$r$  of this point varies as $p(r)\approx (r/R)^{\alpha }$, and when this
vicinity has received $r^{D}$ walkers the local structure up to
length scale $r$ will have been completely regrown.  This requires a number
of walkers added to the cluster given by $\delta N (r/R)^{\alpha }\approx
r^{D}$ leading to $\delta N\approx R^{\alpha }r^{D-\alpha }$.  The argument
is consistent for $\alpha <D$ because it shows that for length scales larger
than $r$ the regrowth threshold will not have been met,  validating our
retention of the hit probability $p(r)\approx (r/R)^{\alpha }$.  From the
point of view of site $X$ significant reorganisation of the cluster happens
first on the smallest scales,  and requires $\delta N\approx R^{\alpha }$
by which point we can expect to have hit site $X$ itself with non-zero
probability.  For $\alpha >D$ the scenario is of screening being dominated
by distant growth and hitting the site is unlikely.  A more detailed
discussion of the screening dynamics is given by Ball and Blunt
\cite{ball90blunt}.

\begin{figure}
\resizebox{\figurewidth}{!}{\includegraphics*{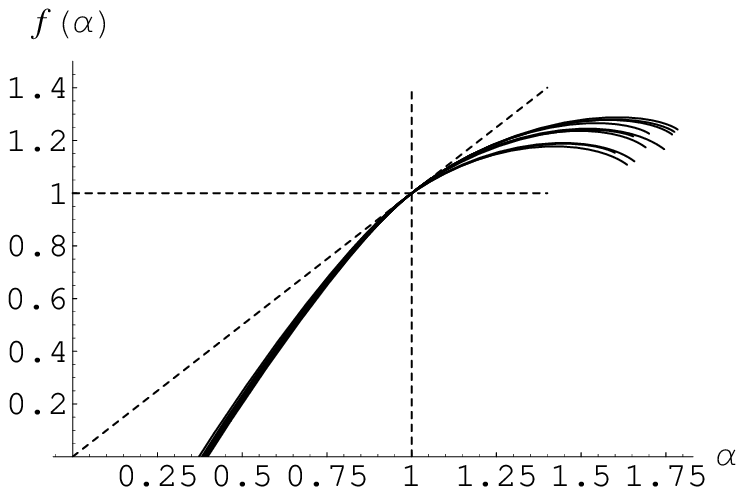}}
\caption{\label{fig:ndep}
Size dependence of the $f(\alpha)$ spectrum of DLA, obtained using first
order ages and clusters grown per Fig.~\ref{fig:mfspec} to different sizes. In
the vicinity $\alpha \lesssim
D$ the curves systematically rise with the range of $N$ used, in support of
our claim that the age method ultimately probes all the way to $\alpha=D$.
The curves are based on
the slope of $\ln Z(q,0,N)$ {\it vs} $\ln N$ across successive factors of
$\sqrt{10}$ in $N$ up to $N=10^7$ (topmost curve).
}
\end{figure}

Figure~\ref{fig:ndep} shows how the $f(\alpha )$ spectrum from first order
ages develops as a function of size up to $N=10^{7}$.  As
expected the
spectrum builds up on the RHS as we increase $N$ consistent with our
expectation that ultimately we can probe as far as $\alpha =D$.
Quantitatively the convergence is poor beyond about $\alpha \simeq 1.5$.

\section{Dielectric Breakdown Model}

\subsection{The random walker model}

Niemeyer, Pietronero and Wiesmann \cite{niemeyer84} introduced the
generalisation of diffusion controlled growth in which the local advance
velocity is set by $\mu ^{\eta }$. In this model $\eta =1$ corresponds to DLA,
while growth with higher values of $\eta $ was proposed to model the evolution
of dielectric breakdown patterns. More recently it has been argued by two of us
\cite{ball02prl,ball03pre} that non-trivial values of $\eta $ can also be
induced when mapping between different types of ultraviolet cutoff, even when
the underlying growth is strictly proportional to flux. The computational
challenge of the DBM is that random walkers sample the harmonic measure
proportional to $\mu $, so for $\eta \neq 1$ an explicit calculation of the
measure appears to be required.  Relaxation methods
\cite{amitrano89,sanchez93} and more recently (for two dimensions) conformal
mapping techniques \cite{hastings01prl} have been used, but the resulting
computations are dramatically more expensive than for simple DLA simulation.

We propose that the following random walker growth model is equivalent to DBM.
We fire random walkers at the cluster, whose first encounter linearly samples
$\mu $. At each encounter we use the $k$'th order age of the site hit to
give the estimate $\mu _{k}=k/a_{k}$ and accordingly advance the growth locally by an
amount proportional to $\delta m=Aa_{k}^{1-\eta }$, which is interpreted as the
mass added to the cluster.  There are two clear constraints, the first is that
the local advance must converge in the mean: $\left\langle a_{k}^{1-\eta
}\right\rangle _{\mu }\propto \mu ^{\eta -1}$, requiring that $k>\eta -1$. The
second is that we desire that the $k$'th order ages should look back no further
than a particle size, so individual advance steps should be bounded by $1/k$.
Na\"{\i}vely this requires $A<1/k$ for $\eta >1$ and $A<N^{\eta -1}$ for $\eta
<1$, where $N$ is the total number of walkers, but we will discuss below how
we can greatly improve on these constraints by adjusting $A$ dynamically.

\subsection{DBM with optimal growth steps}

We now consider allowing the growth step prefactors $A$ to vary as the
growth proceeds. It is convenient to interpret $A=\delta Q$ as the charge
borne by each walker adding to the growth, and the growth is governed by the local advance rate
$\left(\partial r /\partial Q \right)_n=\mu ^{\eta }$ where $Q$ is the cumulative charge
added. It was noted by two of us \cite{ball03pre} that if we
characterise the extremal tips of the growth as having $\mu _{\text{tip}
}\approx R^{-\alpha _{\text{tip}}}$ that the growth law at the tips
trivially leads to $Q\approx R^{1+\eta \alpha _{\text{tip}}}$ as a
generalisation of a standard relation for the fractal dimension of DLA,
and less trivially it is expected that $1+\eta \alpha _{\text{tip}}=\tau
(2+\eta )$ which we will use below to render some expressions less
cumbersome. Within all this framework we are now free to choose the charge
increments per walker $\delta Q$ to vary systematically with growth of a
cluster (but not biased by where each particular walker hits), with the corresponding mass increments given by $\delta m=\delta
Q a_{k}^{1-\eta }$.

We focus on the case $\eta >1$ for which the concern lies with
anomalously low ages which would give large growth steps.  At given $\mu
=R^{-\alpha }$, the probability that a single sampling of the $k$'th age is
less than $a_{k}$ can be found from Eq.~(\ref{pk}) and varies as
$(R^{-\alpha }a_{k})^{k}$ for $\mu a_{k}\ll 1$.  If we grow the cluster by a
significant fraction of its size then the number of walkers sampling this
distribution is given by $(Q/\delta Q) R^{f(\alpha )-\alpha }$ where the
first factor is the total number of walkers added and the second is the
probability of one walker landing on a site with $\mu \approx R^{-\alpha }$.
Therefore the number of times ages below $a_k$ are expected to be sampled is given by
\begin{equation}
\label{eq:max}
\max_{\alpha }\left[ Q/\delta Q R^{f(\alpha )-\alpha }(R^{-\alpha
}a_{k})^{k}\right],
\end{equation}
where we have taken the maximally contributing value of $\alpha$.  To find
the smallest likely value of $a_k$ we then set this expression to unity.
Maximising the $\alpha$ dependence gives $\max_{\alpha }\left[ R^{f(\alpha
)-\alpha }R^{-k\alpha }\right] \approx R^{-\tau (k+1)}$ and it is convenient
to similarly substitute $Q\approx R^{\tau (2+\eta )}$,  leading to expected
minimum age given by
\begin{subequations}
\label{eq:minage}
\begin{equation}
\min \{a_{k}\}\approx \left( R^{\tau (k+1)-\tau (2+\eta )}\delta Q\right)
^{1/k}.
\end{equation}
The above calculation is valid when it yields an increasing function of $R$;
otherwise the scaling assumption in Eq.~(\ref{eq:max}) does not hold, and we
have
\begin{equation}
\min \{a_{k}\} = 1.
\end{equation}
\end{subequations}

\begin{figure*}
\resizebox{0.9 \textwidth}{!}{\includegraphics*{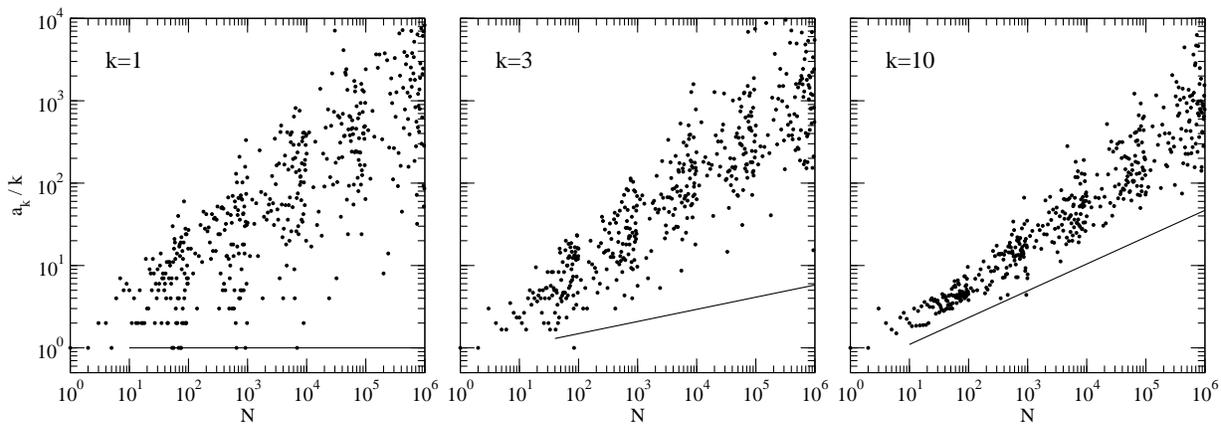}}
\caption{\label{fig:ages}
Scatter plot of the ages of sites hit for DLA: $\eta=1$, $\delta Q=1$.  The
data points are decimated for large $N$ for clarity.  The solid lines
correspond to the expected scaling of minimal ages obtained from Eqns.~(\ref{eq:minage}ab);
these theoretical estimates appear to be suitably cautious.
}
\end{figure*}

We have tested the predicted minimum age by looking at the extremal
ages sampled during DLA cluster growth,  where we have $\delta Q=1$,
$\eta =1$ and $\tau (3)=D$. Figure~\ref{fig:ages} shows scatter plots of all
the ages as a function of cluster growth in terms of $N\approx R^{D}$,  using
ages of different orders and comparing with the result above.

We can now exploit the expected minimum age to set an optimal choice of
$\delta Q$ in DBM growth: we substitute $\min \{a_{k}\}$ into the restriction
that the growth steps be suitably limited $\delta m\ll 1$,  leading to
\begin{equation}
\delta Q\approx R^{(\eta -1)\frac{\tau (k+1)-\tau (2+\eta )}{k+1-\eta }}.
\label{eq:deltaQ}
\end{equation}
Given that we are restricted to $k+1>\eta $, the exponent of $R$ in the
expression for $\delta Q$ increases monotonically with $k$ towards asymptote
$(\eta -1)\alpha _{\text{min}}$ as $k\rightarrow \infty $.

In practice we are inhibited from setting $k$ too large because
$k$-dependent prefactors compete with optimising the power law for finite $R$,
but taking $k=10$ puts us quite close to the limit.  Also, we cannot
use Eq.~(\ref{eq:deltaQ}) explicitly because the values of $\tau(k+1)$ and
$\tau(2+\eta)$ are not known {a priori} for a generic $\eta$.  Instead, armed
with the knowledge that such an asymptotic power law exists, we use the
empirical formula $\delta Q=A_0 k^{\eta-2} N^\beta$ for which the parameters
are chosen such that the constraints are met.  For the parameter values of
Table~\ref{tab:parameters}, used in the simulations below, the condition
$\delta m < 1$ always holds whilst the distance looked back by the $k$'th order
age only exceeds one particle diameter a few times per million walkers and never
exceeds two particle diameters.

Our algorithm is most competitive for $\eta$ near 1, including the important
region $1<\eta<2$ corresponding to surface tension regularization
\cite{ball02prl,ball03pre}, where it can generate cluster masses inaccessible
by other methods.  For too small or large $\eta$ (particularly  $\eta \gtrsim 3$) it
becomes less advantageous.

\begin{table}
\caption{ \label{tab:parameters}
Numerical values of the parameters used to grow walker-DBM clusters.
The local growth is $\delta m = (A_0/k) N^\beta (a_k/k)^{1-\eta}$.
}
\medskip
\begin{ruledtabular}
\begin{tabular}{dddd}
\eta &   k    &  A_0   & \beta  \\ \hline
0.5  &   1    &  0.8   & -0.45   \\
1.5  &   10   &  0.85  &  0.138  \\
2    &   10   &  0.8   &  0.234  \\
3    &   10   &  0.85  &  0.32   \\
\end{tabular}
\end{ruledtabular}
\end{table}

\begin{figure}
\resizebox{\figurewidth}{!}{\includegraphics*{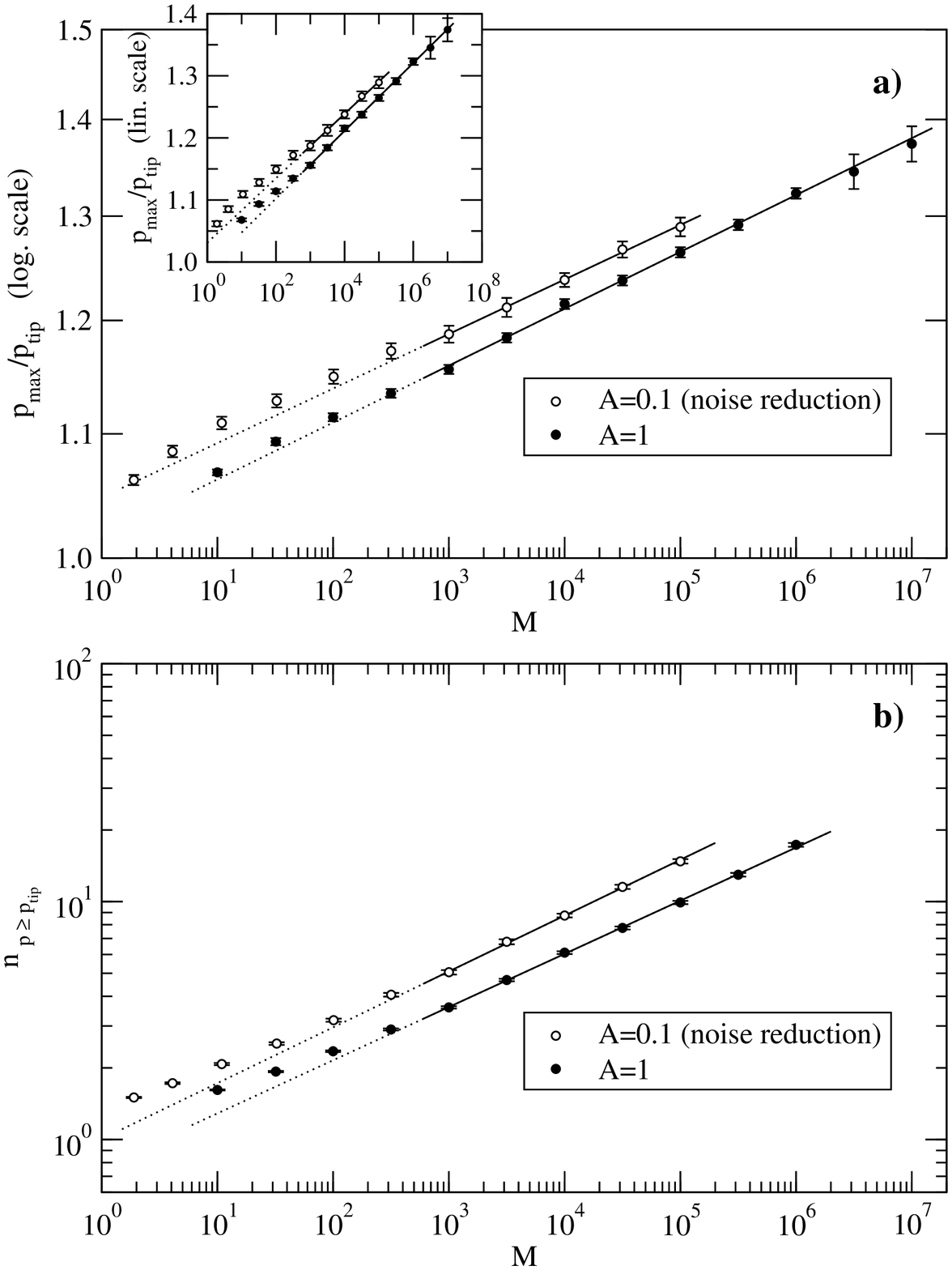}}
\caption{\label{fig:alphamin}
Study of the highest probability growth sites and the tips of DLA clusters.
We launched probe walkers onto a fixed non-growing cluster and recorded the
number of probes landed on each site.  This enables to calculate the growth
probability of each site, including the tip (furthest site from the origin).
\textbf{a)}~The ratio of the growth probabilities of the highest hit rate site
and the tip:  it is a slowly increasing function of the cluster mass $M$.
The best fit curve is a power law with a small exponent:
$p_\text{max}/p_\text{tip} \sim M^{(\alpha_\text{tip}-\alpha_\text{min})/D}$,
with $\alpha_\text{tip}-\alpha_\text{min} = 0.03\pm0.005$. However, one cannot
entirely exclude logarithmic corrections (see \textbf{inset}); in that case
$\alpha_\text{tip} = \alpha_\text{min}$.
\textbf{b)}~The number of sites with higher hit rate than the tip.  This scales
as a power of the radius $R$ with the fractal dimension of tips as exponent:
$n_{p\ge p_\text{tip}} \sim R^{f_\text{tip}} \sim M^{f_\text{tip}/D}$.  The
exponent is larger, clearly distinct from zero: $f_\text{tip} = 0.38\pm 0.03$.
On all plots the two datasets are standard DLA ($A=1$) with $10^4$ to $10^5$
clusters depending on size, and noise reduced DLA ($A=0.1$) with $4200$ to
$10^4$ clusters.  The number of probes was set such that the tip of each
cluster received $2500$ hits.  The solid lines are best fit for the data
points under them; dotted lines are extensions towards excluded data points.
}
\end{figure}

In principle for $k\rightarrow \infty $ the number of walkers required to grow
a cluster is given by
\[
N=\int \frac{dQ}{\delta Q}\approx R^{1+\alpha _{\text{min}}+\eta (\alpha
_{\text{tip}}-\alpha _{\text{min}})}.
\]
The result is familiar in that for $\eta =1$ it recovers the DLA result
(where $N$ is also the cluster mass).  There has been recent controversy
about the precise value of $\alpha _{\text{tip}}-\alpha _{\text{min}}$
through which our method is explicitly sensitive to $\eta $ \cite{jensen03},
although certainly this difference is very small at $\eta =1$,
where $0\le\alpha _{\text{tip}}-\alpha _{\text{min}}\le0.03\pm 0.005$, see
Fig.~\ref{fig:alphamin}.

\subsection{Numerical results}

\begin{figure}
\resizebox{\figurewidth}{!}{\includegraphics*{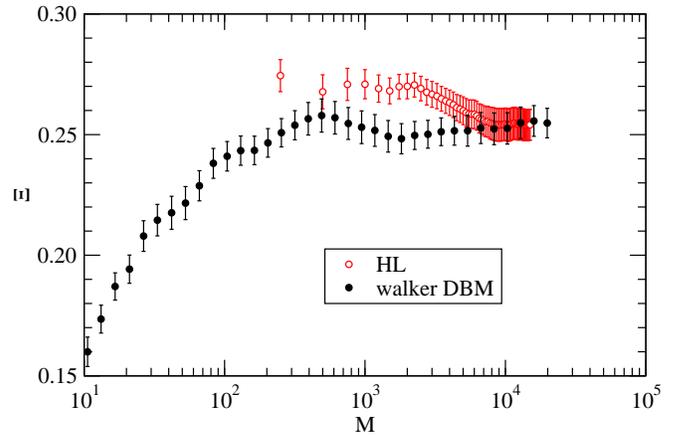}}
\caption{\label{fig:penet}
(Color online) Comparison of the relative penetration depth ~ $\Xi=
\sqrt{\langle r^2\rangle - \langle r\rangle^2} \, / \, \langle r\rangle$ ~ for
the walker DBM and HL methods. For each method the penetration depth was
measured on 30 DBM clusters, $\eta=2$, with 3000 probes on each cluster. Both
methods converge to the same asymptotic value, with the walker-DBM converging
slightly faster.
}
\end{figure}

As a test of the new walker-DBM method, we compared the relative penetration depth
(the growth zone width, normalized by the average deposition radius) with
measurements on clusters grown by the more established HL method, for $\eta=2$.
The asymptotic behavior of the penetration depth can be considered a sensitive
measurement, as for DLA it caused considerable controversy in the past. As seen on
Fig.~\ref{fig:penet}, both methods converge to the same asymptotic value of the
relative penetration depth, and of the two the walker-DBM converges slightly
faster.  This can be understood in terms of the higher effective noise reduction
levels.

\begin{figure}
\resizebox{\figurewidth}{!}{\includegraphics*{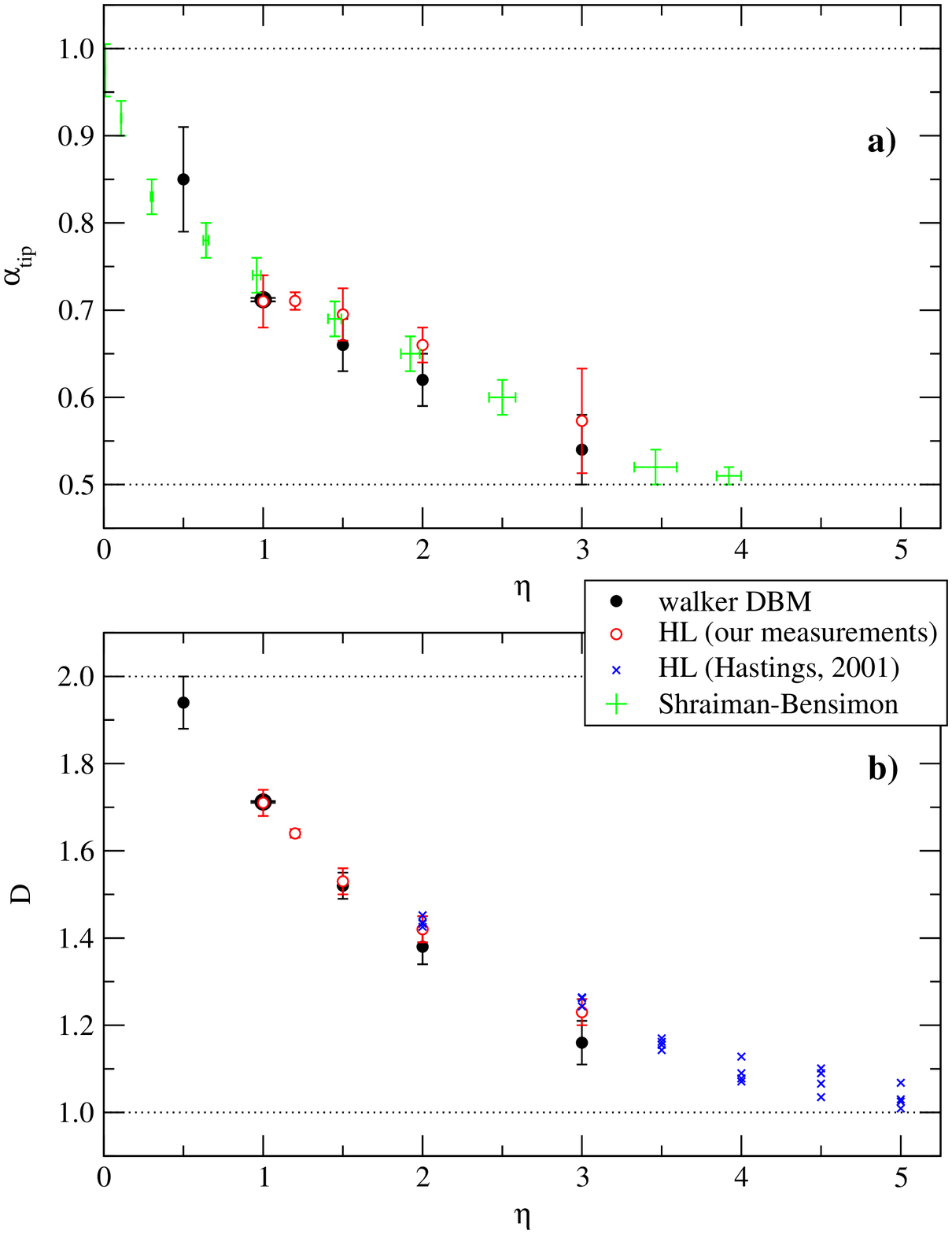}}
\caption{\label{fig:alphadim}
(Color online) \textbf{a)} The tip scaling exponent $\alpha_\text{tip}$ and
\textbf{b)} the fractal dimension $D$ as a function of $\eta$.  The four data
sets plotted are the walker-DBM, our HL measurements, earlier HL measurements
of Ref.~\cite{hastings01prl}, and our Shraiman-Bensimon results.
Because we have to convert the SB results back from
measurements at fixed $\eta_1$ to the corresponding values of
$\eta_0=\eta_1/\alpha$, these points have error bars on both $\alpha$ and
$\eta_0$. The ``walker'' data point for $\eta=1$ is in fact a DLA measurement
(with very small uncertainties) from
Ref.~\cite{somfai03} together with the relation $\alpha_\text{tip}=D-1$. The
dotted lines correspond to the limiting behaviors: dense 2-dimensional growth
($\eta=0$) and non-branching 1-dimensional growth ($\eta\ge\eta_c$).
}
\end{figure}

To probe the scaling, we measured the tip scaling exponent $\alpha_\text{tip}$ and
the fractal dimension $D$ for several values of $\eta$. On Fig.~\ref{fig:alphadim}
we compare these results with measurements obtained with different methods: HL and
by integrating the Shraiman-Bensimon equation; the latter described in more detail
in the next section. The data agree within the stated uncertainties, showing a
smooth monotonic transition between the extremes $\eta=0$ and $\eta\gtrsim\eta_c$,
while below the level of significance point-by-point one can see some differences.
One difference is that for HL, $\alpha_\text{tip}$ shows very little variation
between $1\le\eta\lesssim 2$ (seemingly agreeing with an earlier prediction
\cite{ball02prl,ball03pre}), while the walker-DBM does not show this feature.
Another difference is that for high $\eta$ the walker-DBM yields lower values for
both $\alpha_\text{tip}$ and $D$ than HL.  Although for walker-DBM it was not
practical to obtain values for $\eta>3$, its extrapolation is more consistent with
$\eta_c=4$.

\section{Shraiman-Bensimon Equation}
\label{sec:bs}

Shraiman and Bensimon observed \cite{shraiman84} that for DLA in two
dimensions the evolution of the complex potential reduces to a non-local
problem in one space dimension (plus time).  This is based on the cluster boundary
$z(\theta )$ expressed as a complex function of the imaginary part $\theta$ of
the complex potential, corresponding to cumulative harmonic measure or more
loosely,  charge.  In Refs.~\cite{ball02prl,ball03pre} we adapted this to the
DBM class, arguing that changing from a fixed cutoff scale in physical space
to a fixed cutoff in charge space could be offset by adjustment of the DBM
parameter from $\eta _{0}$ to $\eta _{1}=\alpha \eta _{0}.$ Then in terms of
\[
(-i\partial z/\partial \theta )^{-(1+\eta _{1})/2}=\psi (\theta
,t)=1+\sum_{k=1}^{K}\psi _{k}(t)e^{-ik\theta }
\]
the dynamics for DBM growth along a channel of width $2\pi $ with periodic
boundary conditions reduces to

\begin{eqnarray}
\lefteqn{\frac{d}{dt}\psi _{k}=-k\psi _{k}\sum_{m=0}^{K}\psi _{m}\overline{\psi _{m}}} \nonumber\\%
& &
+\sum_{j=1}^{k}\sum_{m=0}^{K-j}((3+\eta _{1})j-2k)\psi _{k-j}\psi _{j+m}%
\overline{\psi _{m}}
\label{eq:shrb}
\end{eqnarray}

\noindent
with $\psi _{0}=1$. The fixed bandwidth limit $k\leq K$ corresponds
qualitatively to imposing a cutoff of fixed minimal charge on the growing
tips, and inevitably means that the most screened regions of the physical
growth are not tracked.  Scaling arguments lead to the expectation that
$\left\langle \psi _{m}\overline{\psi _{m}}\right\rangle \varpropto
m^{-1+\eta _{1}(1-1/\alpha )}$ for $1\ll m\ll K$,  but give no indication
of how wide a range of $m$ is required to see the scaling behaviour and
hence determine the tip scaling exponent $\alpha $,  and we will see that
this is a major issue below.

The trilinear form of the RHS is efficiently evaluated by Fourier methods
costing of order $K\ln K$ per timestep for the whole system, and in the
results presented here we used $K=1023$ and a simple predictor-corrector
timestepping scheme.  We set the timestep adaptively such that for every
$\psi _{k}$ at each timestep:  \textit{either} the predicted and
corrected updates $\delta \psi _{k}$ agreed within 10\% \textit{or else}
$\psi _{k}$ was being updated by less than 20\%.  As predictor-corrector is
a second order method,  the resulting worst case error is of order 5\%: \
even this sounds generous but as it was applied to the worst case of $1023$
variables the typical precision achieved was very much higher,  and we
chose 20\% on the basis of obtaining results statistically indistinguishable
under refinement.

\begin{figure}
\resizebox{\figurewidth}{!}{
\includegraphics*{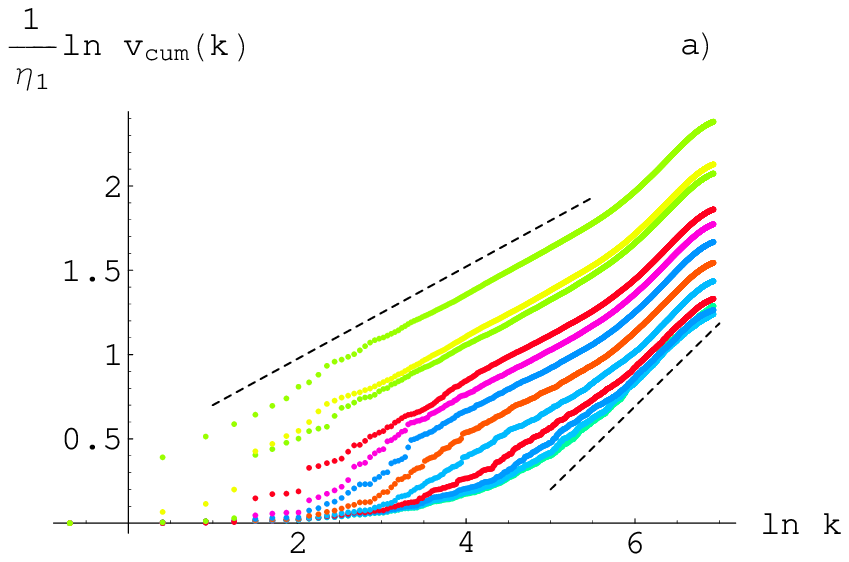}}
\resizebox{\figurewidth}{!}{
\includegraphics*{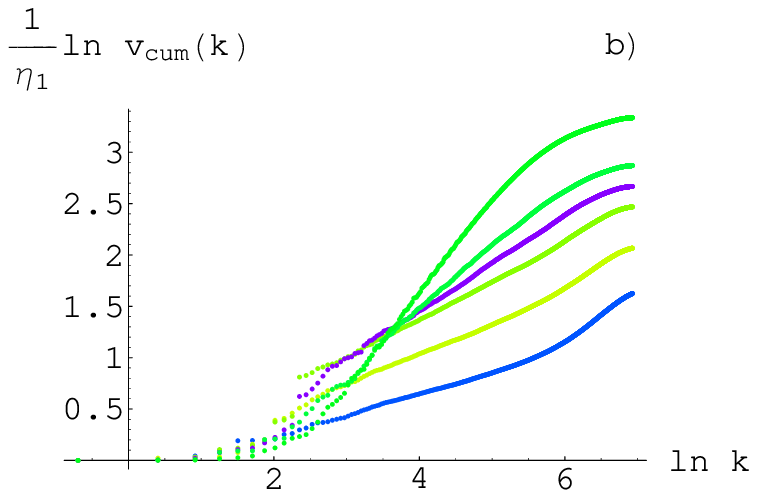}}
\caption{\label{fig:bseqn}
(Color online)
Direct integration of the Bensimon-Shraiman equation (\ref{eq:shrb}) as described in
section \ref{sec:bs} leads to the prediction that
$v_{\rm cum}(k)=\sum_{j<k}\left| \psi _{j}\right| ^{2}$ should vary as
$k^{\eta_1(1-1/\alpha)}$ so the displayed graphs should exhibit slope
$1-1/\alpha$. (Recall that $\eta_1 = \alpha\eta$.)
The upper panel \textbf{a)} shows plots from the same simulation at
$\eta_1=0.5$ for successively later times (bottom to top).  There is characteristic
slope at high $k$ which persists over a limited range, but it is the lower
slope which emerges and eventually dominates the wider range down to the smallest
range of $k$, and which we take to reveal the true tip scaling exponent $\alpha$.
The lower panel \textbf{b)} shows data at large times for $\eta_1=0.3, 0.6, 0.9, 1.2, 1.5, 1.8$ (top to bottom at right).
Whilst the high $k$ slope is insensitive to $\eta_1$ there is a clear
variation of the asymptotic slopes from which we take $\alpha$ in Fig.~\ref{fig:alphadim}(a). }
\end{figure}

Figure \ref{fig:bseqn}(a) shows the observed behaviour of $\left(
\sum_{j<k}\left| \psi _{j}\right| ^{2}\right) ^{1/\eta _{1}}$ vs $k$ with
increasing time for a representative value of $\eta _{1}$. On logarithmic
scales the expectation is a straight line with slope $1-1/\alpha $ and a
difficulty is immediately apparent,  in that two slope behaviour is clearly
observed.  It is clear that it is the slope emerging from lower $k$ values which
predominates at large
times, although different slope is preserved over a limited upper range of $k$. Figure \ref{fig:bseqn}(b) shows results at long times for various $\eta_1$ from which it is
clear that whilst the high $k$ slopes are insensitive to $\eta$, the (we claim)
asymptotic slopes at lower $k$ exhibit a systematic variation.

From the present data the conclusion has to be that the slopes in the lower range
of $k$ give our best estimate of $1-1/\alpha $ and it is these results which are
compared (and agree) with the results by other methods in Fig.~\ref{fig:alphadim}(a).  The
relatively constant slopes at high $k$ are what dominated our earlier numerical
conclusions in \cite{ball02prl,ball03pre} which had a factor of 10 less range in $k$, and appeared to lend
support to the claim that the slope, and hence $\alpha$, might be independent of
$\eta$.  That earlier suggestion is clearly refuted, although we cannot rule out
more surprises from simulations at decades larger $K$.

\section{Conclusions}

We have shown that recurrence times between random walkers provide estimates of the local
harmonic measure which are highly effective for estimating moments and for simulating non-linear growth models.

For the scaling of multifractal moments this technique is limited to the
regions of the growth which are sufficiently active that nearby hits dominate
the screening of a given site, but in that ``active range'' our results for
DLA in two dimensions clearly surpass the quality of previous data and offer
more conclusive support to the theoretical exponent relations of Makarov and
Halsey.  Given that our method is also applicable in higher dimensions, we
believe we have established it as the technique of choice to determine
harmonic measure in the active range.

Our walker-based simulations of the Dielectric Breakdown Model  have greatly
consolidated knowledge of how DBM exponents vary with the nonlinearity
parameter $\eta$ in two dimensions.  In particular we can now be much more
confident of the continuous variation of fractal dimension with $\eta$,
looking at the trend in terms of $\alpha_\text{tip}$ sharpens the issue, and
we can be fairly confident that DBM becomes trivial for  $\eta >\eta_c=4$ as
conjectured theoretically \cite{hastings01pre, hastings01prl, halsey02,
sanchez93}.

The clearest results for the scaling of the DBM model in two dimensions over the full range of $\eta$ now come, ironically,  from our integration of the Shraiman-Bensimon equation.  However unravelling the confusing two slope behaviour which this approach exhibits for the extraction of the exponent $\alpha_\text{tip}$, would not have been prompted without the walker-DBM results.  Overall it is the agreement of both methods with earlier results which establish the definitive picture.

Finally we note that the most crucial role of walker-based DBM will be in three dimensions, where the techniques based on complex analysis have nothing to offer and relatively little is known about the exponent behaviour.  We look forward to exploring this in a subsequent paper.

\begin{acknowledgments}
This research has been supported by the EC under Contract No.
HPMF-CT-2000-00800.
The computing facilities were provided by the Centre for Scientific Computing
of the University of Warwick, with support from the JREI.
We thank Joachim Mathiesen and Anders Levermann for providing us with the data
of Ref.~\cite{jensen02}.
\end{acknowledgments}

\bibliography{dlaref}

\begin{thebibliography}{27}
\expandafter\ifx\csname natexlab\endcsname\relax\def\natexlab#1{#1}\fi
\expandafter\ifx\csname bibnamefont\endcsname\relax
  \def\bibnamefont#1{#1}\fi
\expandafter\ifx\csname bibfnamefont\endcsname\relax
  \def\bibfnamefont#1{#1}\fi
\expandafter\ifx\csname citenamefont\endcsname\relax
  \def\citenamefont#1{#1}\fi
\expandafter\ifx\csname url\endcsname\relax
  \def\url#1{\texttt{#1}}\fi
\expandafter\ifx\csname urlprefix\endcsname\relax\def\urlprefix{URL }\fi
\providecommand{\bibinfo}[2]{#2}
\providecommand{\eprint}[2][]{\url{#2}}

\bibitem[{\citenamefont{Witten and Sander}(1981)}]{witten81}
\bibinfo{author}{\bibfnamefont{T.~A.} \bibnamefont{Witten}} \bibnamefont{and}
  \bibinfo{author}{\bibfnamefont{L.~M.} \bibnamefont{Sander}},
  \bibinfo{journal}{Phys.\ Rev.\ Lett.} \textbf{\bibinfo{volume}{47}},
  \bibinfo{pages}{1400} (\bibinfo{year}{1981}).

\bibitem[{\citenamefont{Ball and Brady}(1985)}]{ball85}
\bibinfo{author}{\bibfnamefont{R.~C.} \bibnamefont{Ball}} \bibnamefont{and}
  \bibinfo{author}{\bibfnamefont{R.~M.} \bibnamefont{Brady}},
  \bibinfo{journal}{J.\ Phys.\ A} \textbf{\bibinfo{volume}{18}},
  \bibinfo{pages}{L809} (\bibinfo{year}{1985}).

\bibitem[{\citenamefont{Tolman and Meakin}(1989)}]{tolman89}
\bibinfo{author}{\bibfnamefont{S.}~\bibnamefont{Tolman}} \bibnamefont{and}
  \bibinfo{author}{\bibfnamefont{P.}~\bibnamefont{Meakin}},
  \bibinfo{journal}{Phys.\ Rev.\ A} \textbf{\bibinfo{volume}{40}},
  \bibinfo{pages}{428} (\bibinfo{year}{1989}).

\bibitem[{\citenamefont{Nittmann et~al.}(1985)\citenamefont{Nittmann, Daccord,
  and Stanley}}]{nittmann85}
\bibinfo{author}{\bibfnamefont{J.}~\bibnamefont{Nittmann}},
  \bibinfo{author}{\bibfnamefont{G.}~\bibnamefont{Daccord}}, \bibnamefont{and}
  \bibinfo{author}{\bibfnamefont{H.~E.} \bibnamefont{Stanley}},
  \bibinfo{journal}{Nature} \textbf{\bibinfo{volume}{314}},
  \bibinfo{pages}{141} (\bibinfo{year}{1985}).

\bibitem[{\citenamefont{Matsushita et~al.}(1984)\citenamefont{Matsushita, Sano,
  Hayakawa, Honjo, and Sawada}}]{matsushita84}
\bibinfo{author}{\bibfnamefont{M.}~\bibnamefont{Matsushita}},
  \bibinfo{author}{\bibfnamefont{M.}~\bibnamefont{Sano}},
  \bibinfo{author}{\bibfnamefont{Y.}~\bibnamefont{Hayakawa}},
  \bibinfo{author}{\bibfnamefont{H.}~\bibnamefont{Honjo}}, \bibnamefont{and}
  \bibinfo{author}{\bibfnamefont{Y.}~\bibnamefont{Sawada}},
  \bibinfo{journal}{Phys.\ Rev.\ Lett.} \textbf{\bibinfo{volume}{53}},
  \bibinfo{pages}{286} (\bibinfo{year}{1984}).

\bibitem[{\citenamefont{Brady and Ball}(1984)}]{brady84}
\bibinfo{author}{\bibfnamefont{R.~M.} \bibnamefont{Brady}} \bibnamefont{and}
  \bibinfo{author}{\bibfnamefont{R.~C.} \bibnamefont{Ball}},
  \bibinfo{journal}{Nature} \textbf{\bibinfo{volume}{309}},
  \bibinfo{pages}{225} (\bibinfo{year}{1984}).

\bibitem[{\citenamefont{Somfai et~al.}(1999)\citenamefont{Somfai, Sander, and
  Ball}}]{somfai99}
\bibinfo{author}{\bibfnamefont{E.}~\bibnamefont{Somfai}},
  \bibinfo{author}{\bibfnamefont{L.~M.} \bibnamefont{Sander}},
  \bibnamefont{and} \bibinfo{author}{\bibfnamefont{R.~C.} \bibnamefont{Ball}},
  \bibinfo{journal}{Phys.\ Rev.\ Lett.} \textbf{\bibinfo{volume}{83}},
  \bibinfo{pages}{5523} (\bibinfo{year}{1999}), \eprint{cond-mat/9909040}.

\bibitem[{\citenamefont{Amitrano}(1989)}]{amitrano89}
\bibinfo{author}{\bibfnamefont{C.}~\bibnamefont{Amitrano}},
  \bibinfo{journal}{Phys.\ Rev.\ A} \textbf{\bibinfo{volume}{39}},
  \bibinfo{pages}{6618} (\bibinfo{year}{1989}).

\bibitem[{\citenamefont{Ball and Blunt}(1990)}]{ball90blunt}
\bibinfo{author}{\bibfnamefont{R.}~\bibnamefont{Ball}} \bibnamefont{and}
  \bibinfo{author}{\bibfnamefont{M.}~\bibnamefont{Blunt}},
  \bibinfo{journal}{Phys.\ Rev.\ A} \textbf{\bibinfo{volume}{41}},
  \bibinfo{pages}{582} (\bibinfo{year}{1990}).

\bibitem[{\citenamefont{Jensen et~al.}(2003)\citenamefont{Jensen, Mathiesen,
  and Procaccia}}]{jensen03}
\bibinfo{author}{\bibfnamefont{M.~H.} \bibnamefont{Jensen}},
  \bibinfo{author}{\bibfnamefont{J.}~\bibnamefont{Mathiesen}},
  \bibnamefont{and}
  \bibinfo{author}{\bibfnamefont{I.}~\bibnamefont{Procaccia}},
  \bibinfo{journal}{Phys.\ Rev.\ E} \textbf{\bibinfo{volume}{67}},
  \bibinfo{pages}{042402} (\bibinfo{year}{2003}).

\bibitem[{\citenamefont{Halsey et~al.}(1986{\natexlab{a}})\citenamefont{Halsey,
  Meakin, and Procaccia}}]{halsey86prl}
\bibinfo{author}{\bibfnamefont{T.~C.} \bibnamefont{Halsey}},
  \bibinfo{author}{\bibfnamefont{P.}~\bibnamefont{Meakin}}, \bibnamefont{and}
  \bibinfo{author}{\bibfnamefont{I.}~\bibnamefont{Procaccia}},
  \bibinfo{journal}{Phys.\ Rev.\ Lett.} \textbf{\bibinfo{volume}{56}},
  \bibinfo{pages}{854} (\bibinfo{year}{1986}{\natexlab{a}}).

\bibitem[{\citenamefont{Niemeyer et~al.}(1984)\citenamefont{Niemeyer,
  Pietronero, and Wiesmann}}]{niemeyer84}
\bibinfo{author}{\bibfnamefont{L.}~\bibnamefont{Niemeyer}},
  \bibinfo{author}{\bibfnamefont{L.}~\bibnamefont{Pietronero}},
  \bibnamefont{and} \bibinfo{author}{\bibfnamefont{H.~J.}
  \bibnamefont{Wiesmann}}, \bibinfo{journal}{Phys.\ Rev.\ Lett.}
  \textbf{\bibinfo{volume}{52}}, \bibinfo{pages}{1033} (\bibinfo{year}{1984}).

\bibitem[{\citenamefont{Ball and Somfai}(2002)}]{ball02prl}
\bibinfo{author}{\bibfnamefont{R.~C.} \bibnamefont{Ball}} \bibnamefont{and}
  \bibinfo{author}{\bibfnamefont{E.}~\bibnamefont{Somfai}},
  \bibinfo{journal}{Phys.\ Rev.\ Lett.} \textbf{\bibinfo{volume}{89}},
  \bibinfo{pages}{135503} (\bibinfo{year}{2002}), \eprint{cond-mat/0205660}.

\bibitem[{\citenamefont{Ball and Somfai}(2003)}]{ball03pre}
\bibinfo{author}{\bibfnamefont{R.~C.} \bibnamefont{Ball}} \bibnamefont{and}
  \bibinfo{author}{\bibfnamefont{E.}~\bibnamefont{Somfai}},
  \bibinfo{journal}{Phys.\ Rev.\ E} \textbf{\bibinfo{volume}{67}},
  \bibinfo{pages}{021401} (\bibinfo{year}{2003}), \eprint{cond-mat/0210598}.

\bibitem[{\citenamefont{Ball et~al.}(2002)\citenamefont{Ball, Bowler, Sander,
  and Somfai}}]{ball02pre}
\bibinfo{author}{\bibfnamefont{R.~C.} \bibnamefont{Ball}},
  \bibinfo{author}{\bibfnamefont{N.~E.} \bibnamefont{Bowler}},
  \bibinfo{author}{\bibfnamefont{L.~M.} \bibnamefont{Sander}},
  \bibnamefont{and} \bibinfo{author}{\bibfnamefont{E.}~\bibnamefont{Somfai}},
  \bibinfo{journal}{Phys.\ Rev.\ E} \textbf{\bibinfo{volume}{66}},
  \bibinfo{pages}{026109} (\bibinfo{year}{2002}), \eprint{cond-mat/0108252}.

\bibitem[{\citenamefont{Hastings}(2001{\natexlab{a}})}]{hastings01pre}
\bibinfo{author}{\bibfnamefont{M.~B.} \bibnamefont{Hastings}},
  \bibinfo{journal}{Phys.\ Rev.\ E} \textbf{\bibinfo{volume}{64}},
  \bibinfo{pages}{046104} (\bibinfo{year}{2001}{\natexlab{a}}),
  \eprint{cond-mat/0104344}.

\bibitem[{\citenamefont{Hastings}(2001{\natexlab{b}})}]{hastings01prl}
\bibinfo{author}{\bibfnamefont{M.~B.} \bibnamefont{Hastings}},
  \bibinfo{journal}{Phys.\ Rev.\ Lett.} \textbf{\bibinfo{volume}{87}},
  \bibinfo{pages}{175502} (\bibinfo{year}{2001}{\natexlab{b}}),
  \eprint{cond-mat/0103312}.

\bibitem[{\citenamefont{Halsey}(2002)}]{halsey02}
\bibinfo{author}{\bibfnamefont{T.~C.} \bibnamefont{Halsey}},
  \bibinfo{journal}{Phys.\ Rev.\ E} \textbf{\bibinfo{volume}{65}},
  \bibinfo{pages}{021104} (\bibinfo{year}{2002}), \eprint{cond-mat/0105047}.

\bibitem[{\citenamefont{Sanchez et~al.}(1993)\citenamefont{Sanchez, Guinea,
  Sander, Hakim, and Louis}}]{sanchez93}
\bibinfo{author}{\bibfnamefont{A.}~\bibnamefont{Sanchez}},
  \bibinfo{author}{\bibfnamefont{F.}~\bibnamefont{Guinea}},
  \bibinfo{author}{\bibfnamefont{L.~M.} \bibnamefont{Sander}},
  \bibinfo{author}{\bibfnamefont{V.}~\bibnamefont{Hakim}}, \bibnamefont{and}
  \bibinfo{author}{\bibfnamefont{E.}~\bibnamefont{Louis}},
  \bibinfo{journal}{Phys.\ Rev.\ E} \textbf{\bibinfo{volume}{48}},
  \bibinfo{pages}{1296} (\bibinfo{year}{1993}).

\bibitem[{\citenamefont{Hastings and Levitov}(1998)}]{hastings98}
\bibinfo{author}{\bibfnamefont{M.~B.} \bibnamefont{Hastings}} \bibnamefont{and}
  \bibinfo{author}{\bibfnamefont{L.~S.} \bibnamefont{Levitov}},
  \bibinfo{journal}{Physica D} \textbf{\bibinfo{volume}{116}},
  \bibinfo{pages}{244} (\bibinfo{year}{1998}).

\bibitem[{\citenamefont{Shraiman and Bensimon}(1984)}]{shraiman84}
\bibinfo{author}{\bibfnamefont{B.}~\bibnamefont{Shraiman}} \bibnamefont{and}
  \bibinfo{author}{\bibfnamefont{D.}~\bibnamefont{Bensimon}},
  \bibinfo{journal}{Phys.\ Rev.\ A} \textbf{\bibinfo{volume}{30}},
  \bibinfo{pages}{2840} (\bibinfo{year}{1984}).

\bibitem[{\citenamefont{Halsey et~al.}(1986{\natexlab{b}})\citenamefont{Halsey,
  Jensen, Kadanoff, Procaccia, and Shraiman}}]{halsey86pra}
\bibinfo{author}{\bibfnamefont{T.~C.} \bibnamefont{Halsey}},
  \bibinfo{author}{\bibfnamefont{M.~H.} \bibnamefont{Jensen}},
  \bibinfo{author}{\bibfnamefont{L.~P.} \bibnamefont{Kadanoff}},
  \bibinfo{author}{\bibfnamefont{I.}~\bibnamefont{Procaccia}},
  \bibnamefont{and} \bibinfo{author}{\bibfnamefont{B.~I.}
  \bibnamefont{Shraiman}}, \bibinfo{journal}{Phys.\ Rev.\ A}
  \textbf{\bibinfo{volume}{33}}, \bibinfo{pages}{1141}
  (\bibinfo{year}{1986}{\natexlab{b}}).

\bibitem[{\citenamefont{Halsey}(1987)}]{halsey87}
\bibinfo{author}{\bibfnamefont{T.~C.} \bibnamefont{Halsey}},
  \bibinfo{journal}{Phys.\ Rev.\ Lett.} \textbf{\bibinfo{volume}{59}},
  \bibinfo{pages}{2067} (\bibinfo{year}{1987}).

\bibitem[{\citenamefont{Ball and Spivack}(1990)}]{ball90}
\bibinfo{author}{\bibfnamefont{R.~C.} \bibnamefont{Ball}} \bibnamefont{and}
  \bibinfo{author}{\bibfnamefont{O.~R.} \bibnamefont{Spivack}},
  \bibinfo{journal}{J.\ Phys.\ A} \textbf{\bibinfo{volume}{23}},
  \bibinfo{pages}{5295} (\bibinfo{year}{1990}).

\bibitem[{\citenamefont{Jensen et~al.}(2002)\citenamefont{Jensen, Levermann,
  Mathiesen, and Procaccia}}]{jensen02}
\bibinfo{author}{\bibfnamefont{M.~H.} \bibnamefont{Jensen}},
  \bibinfo{author}{\bibfnamefont{A.}~\bibnamefont{Levermann}},
  \bibinfo{author}{\bibfnamefont{J.}~\bibnamefont{Mathiesen}},
  \bibnamefont{and}
  \bibinfo{author}{\bibfnamefont{I.}~\bibnamefont{Procaccia}},
  \bibinfo{journal}{Phys.\ Rev.\ E} \textbf{\bibinfo{volume}{65}},
  \bibinfo{pages}{046109} (\bibinfo{year}{2002}), \eprint{cond-mat/0110203}.

\bibitem[{\citenamefont{Makarov}(1985)}]{makarov85}
\bibinfo{author}{\bibfnamefont{N.~G.} \bibnamefont{Makarov}},
  \bibinfo{journal}{P.\ Lond.\ Math.\ Soc.} \textbf{\bibinfo{volume}{51}},
  \bibinfo{pages}{369} (\bibinfo{year}{1985}).

\bibitem[{\citenamefont{Somfai et~al.}(2003)\citenamefont{Somfai, Ball, DeVita,
  and Sander}}]{somfai03}
\bibinfo{author}{\bibfnamefont{E.}~\bibnamefont{Somfai}},
  \bibinfo{author}{\bibfnamefont{R.~C.} \bibnamefont{Ball}},
  \bibinfo{author}{\bibfnamefont{J.~P.} \bibnamefont{DeVita}},
  \bibnamefont{and} \bibinfo{author}{\bibfnamefont{L.~M.}
  \bibnamefont{Sander}}, \bibinfo{journal}{Phys.\ Rev.\ E}
  \textbf{\bibinfo{volume}{68}}, \bibinfo{pages}{020401(R)}
  (\bibinfo{year}{2003}), \eprint{cond-mat/0304458v2}.

\end{thebibliography}

\end{document}